# On the analysis of saturated pressure to detect fatigue

Marcos Faundez-Zanuy[1][0000-0003-0605-1282] Josep Lopez-Xarbau[2][0000-0001-5138-8947] Moises Diaz[2][0000-0003-3878-3867] Manuel Garnacho-Castaño[3][0000-0001-5732-7854]

[1] Tecnocampus Universitat Pompeu Fabra, Avda. Ernest Lluch 32, 08302 Mataró, Spain
[2] iDeTIC, Universidad de Las Palmas de Gran Canaria, Juan de Quesada, 30, 35001, Las Palmas de Gran Canaria, Spain
[3] DAFNiS group. Sant Joan de Déu Sant Boi, Doctor Antoni Pujadas, 42, 08830, Sant Boi de Llobregat Spain
`faundez@tecnocampus.cat`

**Abstract.** This paper examines the saturation of pressure signals during various handwriting tasks, including drawings, cursive text, capital words text, and signature, under different levels of fatigue. Experimental results demonstrate a significant rise in the proportion of saturated samples following strenuous exercise in tasks performed without resting wrist. The analysis of saturation highlights significant differences when comparing the results to the baseline situation and strenuous fatigue.

## 1 Introduction

Online handwriting acquisition using digitizing tablets, such as WACOM intuos, has a wide range of applications from e-security to e-health (Faundez-Zanuy et al., 2020). Depending on the specific application, certain tasks may be more suitable than the others, with greater variability in tasks when dealing with e-health applications (Faundez-Zanuy et al., 2021). In all the cases, the writer is required to perform a task such as his own signature, copying a drawing, performing a repetitive task, such as concentric circles, etc. While these tasks may not be specially challenging for healthy individuals, those with conditions like dementia may find them impossible to perform correctly. This inability is used as a biomarker, which is an indicator of pathology. The inability to perform the task is usually detected by incorrect or missing parts of the strokes, tremors, and other factors that mainly focus on the spatial x and y coordinates, with automated analysis replicating the visual inspection performed by health experts, such as in the pentagon drawing test of the Folstein's mini-mental (Folstein et al., 1975). This approach has been successfully applied to analyze prevalent pathologies such as Parkinson disease (Drotár et al., 2016) and Alzheimer disease (Garre-Olmo et al., 2017). In the e-security applications, the challenge is to differentiate a genuine signature from an impostor one. This is a challenging task as impostors can replicate dynamic information, such as spatial coordinates, angles and pressure, from a genuine signature. Although the connection between fatigue and handwriting could be under debate, a fatigued person has poor coordination of the movements and fine motor control skills. It can be manifested in the handwriting in diverse characteristics like the shape of the letters, speed, uneven spacing, and so on. In this paper, we propose a more in-depth analysis of pressure signals and their variation under different levels of fatigue, using an existing database known as Tecnocampus fatigue database (Garnacho-Castaño et al., 2020).

Fatigue can severely affect sport and professional activity performance, including systems operated by brain–computer interface (BCI), which has been an active research field. In [11], the authors studied BCI systems and investigated how fatigue changes with BCI usage and reported the effect of fatigue on signal quality, analyzing five hours of BCI usage. In [12], the authors investigated the effects of three mental states: Fatigue, frustration, and attention on BCI performance. They found that the relationships among these variables were complex, rather than monotonic, probably due to a poorly induced fatigue, which was not assessed by any other mean. In [13], the authors proposed a novel signal-processing chain inspired from BCI computing to detect mental fatigue.

## 1.1 Handwriting Analysis in Fatigue Condition by Calligraphic Experts

Some works based on visual inspection by calligraphic experts reported the effects of fatigue in offline handwriting, where offline is referred to as the case where the user executes a handwriting task and, later on, the task is analyzed. The book [14] classifies the human stress into two forms: Emotional stress and physical stress. Signatures can be executed by a fatigued writer. For instance, when attending a physical fitness center or signing a receipt. However, it is not usual on more formal documents because of the special circumstances of this act. In these cases it should be enough time to recover from fatigue before writing. This can explain the small number of studies dealing with fatigue and handwriting. First studies were published in [15] (p. 94) and [16] (p. 297), but they are considered as one among several factors that may affect handwriting. In [17] (p. 92) the author reported on the changes in the writing of one subject writing a sentence of eighteen words after having run up four floors of stairs. His findings are summarized in [14]. The author reports changes similar to those produced by intoxication, especially an increase in lateral expansion but without an apparent increase in height. In [18] the authors present a study of 30 writers affected by extreme and moderate states of fatigue and fatigue localized to the writers' forearm. The study evaluated healthy males of very similar age (early twenties) who were asked to write a modified version of the London Letter under four different test conditions. The author observed an increase in vertical height (more than 90% of the analyzed cases) in both lowercase and uppercase letters, without a significant change in proportions or relative heights and an increase in letter width or lateral expansion (77% of the subjects). Considering the spacing between words, the author found both phenomena, expansion and contraction (50% of the subjects), but noted that whichever tendency was exhibited it remained consistent. Slope, speed, rhythm, or fluency habits were not significantly affected. Only minor deterioration was appreciated in writing quality, which tended to produce a scrawl and exhibit less care. In only one case there was greater pen pressure displayed. No evidence of tremor was found. When writing under fatigue, fewer patchings and overwritings occurred. Minute movements tended to be enlarged, but fundamental change to most writing habits was not found. There was some propensity to commit spelling errors, to abbreviate, and to omit punctuation and diacritics ("i" dots). There was no apparent difference between the effects of general body fatigue and forearm fatigue. In either case, however, the severity of the fatigue produced some difference. Roulston's data are fully reported in his work and, while the effects of fatigue cannot be denied, Harrison's statement that "…Fatigue and a poor state of health can have a most deleterious effect upon handwriting…" may be an exaggeration of the condition. In [19], the author presents a study of 21 high school students endeavored to assess whether there was a degree of impairment to a person's writing that would correlate with the pulse rate of the heart under different levels of exertion. This paper studied further what the nature of the impairment would be and whether the author of stress-impaired writing could be correctly identified. The author found that physical stress, producing abnormally high pulse rates, did affect the individual's writing performance, but the reading of pulse rates can only be considered an indicator of the level of stress being experienced by the subject. The author of this study could not conclude that pulse rates are fully responsible for the degree of impairment, although they could be a contributing factor. Impairment of the writing in this case was judged on the strength of the following: (1) Deterioration in letter formation, coupled with overwriting and corrections. (2) An increase in lateral expansion, especially of the spacing between letters, and a frequent misjudgment of the length of words at the ends of lines. (3) A tendency to write larger. (4) A reduction in the speed of writing, accompanied by an inconsistency in point load (pen pressure). (5) A failure to maintain good alignment or a proper baseline. (6) A general failure in writing quality and greater carelessness. A set of 15 experienced writing examiners was able to assign correctly the writing samples of each subject (100 percent accuracy) affected by different levels of stress, despite the impairments evidenced in the writings. These results were consistent with those published in [18]. The author presented an interesting closing remark: The impairment of writing produced by physical stress (after running various distances) was generally similar to, but not as pronounced as, the impairment due to alcohol ingestion. Thus, extreme fatigue has some effect on the control of the writing instrument, which tends to increase the expansion of the writing both vertically and horizontally. This expansion may be noted in the enlargement of the more minute movements of the writing process and suggests a trend in the writing toward a scrawl. The effects were of relatively short duration and the writing returned to normal when the body had enough time to recover its energy.

These previous-reported studies, found in [14], are based on visual inspection by calligraphic experts rather than computer analyzed documents. Thus, they can be affected by subjectivity of the human examiner.

Some other relevant works deal with the comparison of the effect of fatigue in people affected by movements disorders and control populations [10], the effect of fatigue in children [15] and the relevance of fatigue on forensic assessment [16].

In [10] a comparison of prefatigue and fatigue handwriting conditions between movement disorder patients and normal controls was conducted of online recordings and offline samples. The offline analysis showed differences due to fatigue. The differences were more pronounced in certain movement-disorder patients (e.g., printing, size, tremor, or baseline) than in the control subjects (e.g., expansion). In the online analysis, the patients were slower and more variable than the controls. Certain patients exhibited increased speed and decreased relative pen down duration during the fatigue condition. Individual responses to fatigue have relevance in the assessment of the range of variation within individuals. Fatigue increases variability in motor-disordered handwriting more severely than in healthy handwriting.

In [15] the authors conclude that children with both poor and good handwriting perform more poorly after writing long texts. Although both groups were influenced by the fatigue situation, poor handwriters still scored lower than the good handwriters in both conditions (fatigue and non-fatigue), on most variables.

The study [16] found that writers with good writing speed and above-average writing skill completed the entire test without important departure from their normal handwriting characteristics. The writers with lower speed and writing skills departed rapidly from their original characteristics as soon as they attempted to improve their writing speed. The result was a rapid tiring of writers with the degeneration of handwriting characteristics to the extent that it may impair its forensic comparison. Among the handwriting features most likely to be modified by fatigue were the following: departure from writing line, increase in absolute size of writing, increase in the relative size of letters, increase in the length of lower and/or upper strokes, increase in the lateral spacing between letters and words, and a general increase in the speed of writing.

## 2    Methodology

### 2.1    Database

In this study we used the Tecnocampus fatigue database (Garnacho-Castaño et al., 2020), which comprises data from 21 healthy male subjects who completed nine different handwritten tasks:
- task 1: Folstein's pentagon copying
- task 2: house drawing copying
- task 3: Archimedes spiral drawing
- task 4: signature
- task 5: repeated concentric circles
- task 6: words in capital letters copying
- task 7: cursive sentence copying
- task 8: signature (same than task 4)
- task 9: spring drawing

These nine tasks are performed under different levels of induced fatigue, and acquired in five different sessions (S1, …, S5). Fatigue was induced by means of a set of physical exercises in young sportive people.

The database was initially developed by the authors for the purpose of detecting fatigue and studying its impact on e-security biometric recognition using signatures and capitalized text (Sesa-Nogueras et al., 2021). The database is not publicly available due to regulations in the data protection law. Figure 1 shows the nine acquired tasks for one user and during one out five acquisition session.

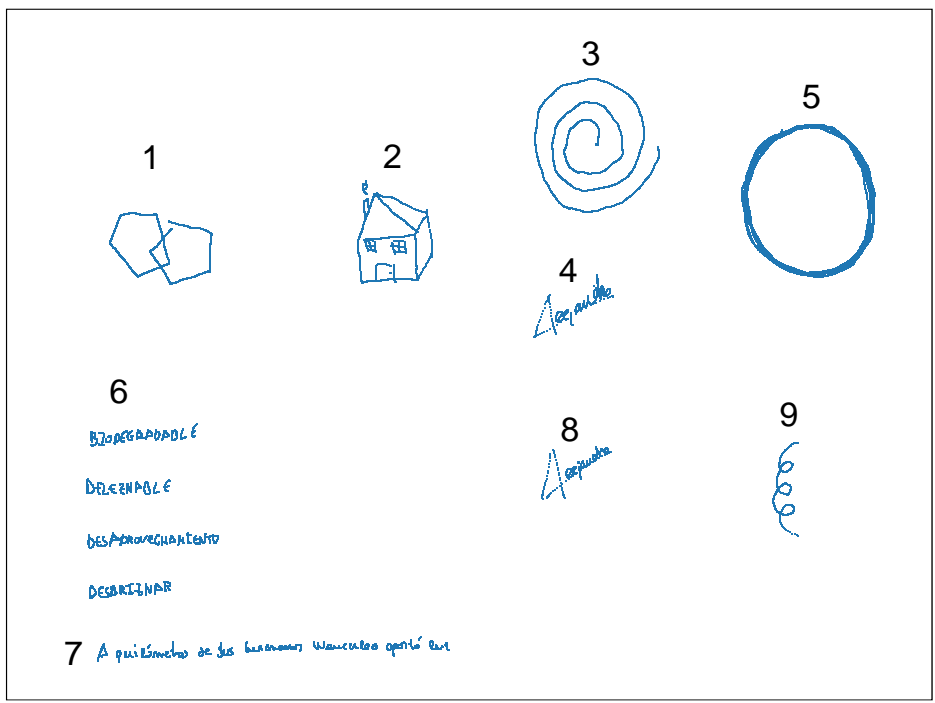

**Fig. 1.** Graphical example of the nine tasks executed by a control enrolled in the Tecnocampus-fatigue database during one out five sessions.

As we can see, the signature was acquired twice per session (tasks 4 and 8). Therefore, there were 8 different tasks in the database.

The sessions are summarized in Figure 2. A complete explanation of the differences between sessions can be found in (Garnacho-Castaño et al., 2020).

## 2.2   Feature extraction

In this paper, we propose a novel feature for handwriting analysis: the percentage of saturated samples in a specific task. This feature can be defined by the equation (1).

$$saturated = \frac{1}{n}\sum_{i=1}^{n}[pressure\_signal_i \geq sat\_level] \tag{1}$$

where: *saturated* contains the proportion of saturated values in a handwriting, *pressure_signal* is the input pressure signal vector, *sat_level* is the given saturation level (depends on tablet model) and *n* is the length of the input pressure signal vector.

Equation (1) is calculated using the MATLAB function shown in Figure 3, where pressure_signal is the pressure vector of the handwritten task and sat_level is the maximum pressure level (i.e. saturation level). For the Wacom tablet used in Tecnocampus fatigue database, sat_level =1023. This saturation level is equivalent to 45 Newton/mm$^2$ following our previous work (Faundez-Zanuy et al., 2021).

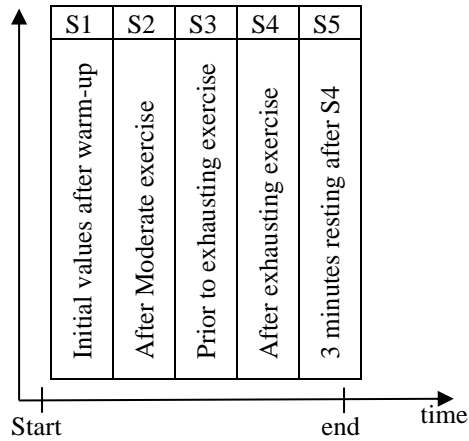

**Fig. 2.** Summary of the five acquisition sessions, where fatigue corresponds to session 4 and 5.

```
function [saturated]=saturation(pressure_signal,sat_level)
   [i]=find(pressure_signal>=sat_level);
   saturated=length(i)/length(pressure_signal);
end
```

**Fig. 3.** Script of MATLAB to count the percentage of saturated samples on the pressure signal of a specific task

## 3     Experimental results

The dots in Figure 4 shows the percentage of saturated samples for each session and task. We can observe two groups of tasks based on the variation between sessions:
a) High variation: There is a variation in the number of saturated samples between the fatigue and rest situation that can be up to five times higher for the pentagon and house, Archimedean spiral and concentric loop copy tasks. These four tasks are marked with bold solid lines in Figure 4.
b) Low variation: There is almost no difference for the tasks of signature, capital letters, spring drawing, and cursive text.

Figure 5 represents the mean pressure for each task and session. From Figure 5, it can be observed that:

- The mean pressure of loops and Archimedes spiral is larger than the mean pressure of the other tasks.
- Similar values of mean pressure are seen in different sessions.
- There is slightly higher pressure in session number 4 than in the other ones for the following tasks: concentric loops, Archimedes spiral, pentagon and house copying test.

Comparing Figures 4 and 5, it can be observed that the percentage of saturated samples exhibits higher sensibility to fatigue than the mean pressure value. As such, the percentage of saturated samples can be considered a potential feature with sensibility to detect fatigue in some specific tasks.

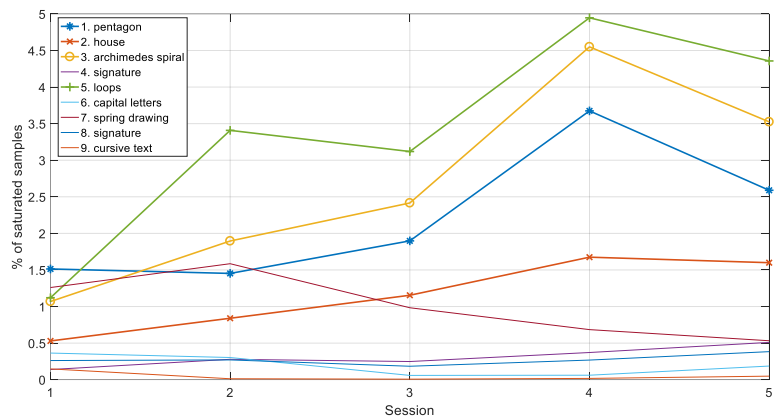

**Fig. 4.** Percentage of saturated samples for each task and session, averaging all the users.

Table 1 shows the standard deviation for the mean pressure values depicted in fig. 5.

**Table 1.** Standard deviation (std) for mean pressure vales depicted in figure 5.

| session | T1 | T2 | T3 | T4 | T5 | T6 | T7 | T8 | T9 |
|---|---|---|---|---|---|---|---|---|---|
| 1 | 242 | 136 | 147 | 128 | 171 | 106 | 209 | 124 | 126 |
| 2 | 211 | 154 | 153 | 152 | 189 | 107 | 212 | 145 | 121 |
| 3 | 240 | 164 | 175 | 174 | 190 | 107 | 206 | 135 | 120 |
| 4 | 239 | 178 | 181 | 169 | 200 | 97 | 221 | 137 | 124 |
| 5 | 237 | 187 | 168 | 150 | 192 | 112 | 216 | 145 | 131 |

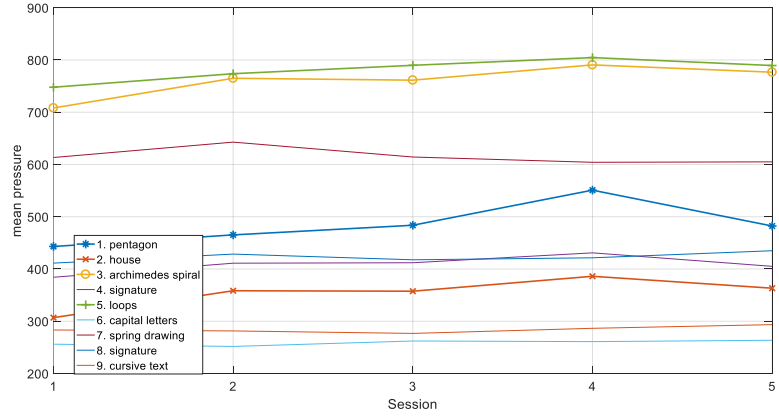

**Fig. 5.** Mean pressure for each task and session, averaging all the users.

In order to quantitatively analyze the differences under fatigue, we present a couple of figures. Figure 6 shows the y-coordinate and pressure of user number 12 for task 5 (concentric loops) in session 1 and 4. Figure 7 shows the speed in x and y coordinates for this same user, task and sessions. In this case we have computed the speed by means of equation 2, where *f* is the feature (x or y-coordinate) and *i* is the sample index.

$$\dot{f}_i = \frac{f_i(l+1) - f_i(l)}{1} = f_i(l+1) - f_i(l) \tag{2}$$

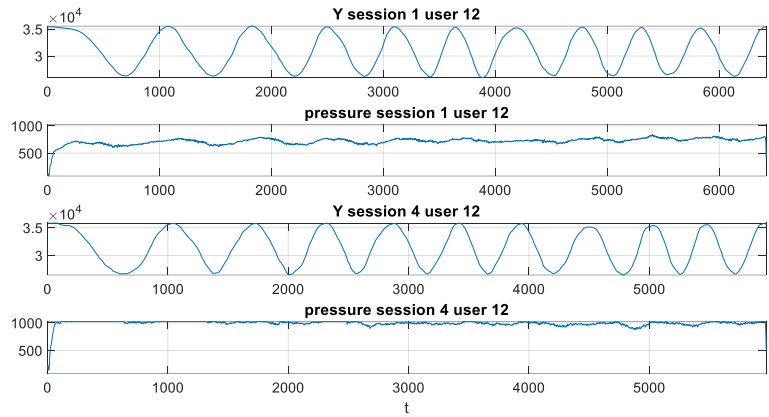

**Fig. 6.** Example of y-coordinate and p signal for loops for the user 12 in session 1 (top) and 4 (bottom).

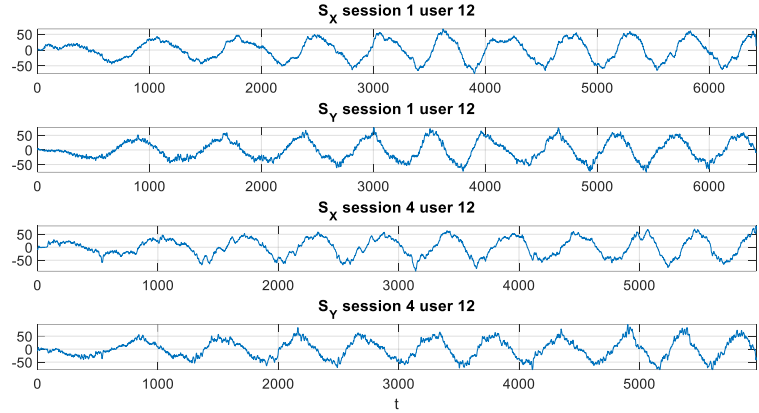

**Fig. 7.** Speed of x and y-coordinates for the user 12 in session 1 (top) and 4 (bottom).

Equation (2) represents an approximation for derivative of a discrete signal, obtained from classical derivative equation for a continuous signal (3).

$$\dot{f}_i = \lim_{h \to 0} \frac{f_i(l+h) - f_i(l)}{h} \tag{3}$$

Another approximation for first derivative is also possible, and has been studied in our recent paper (Faundez-Zanuy & Diaz 2023).

Figure 6 reveals the saturation phenomenon in session 4 (saturation level is 1023). On the other hand, in figure 7 we observe that the dynamics of x and y-coordinate are mainly the same, without evident variations in instantaneous speed.

The tasks that are most affected by fatigue are concentric loops and Archimedes spiral. It is worth pointing out that these tasks require that the user does not rest the wrist on the tablet surface. This implies more difficult pressure control, which seems indeed affected by fatigue. For the pentagon and house drawing tests, although some parts can be performed with wrist resting, wrist up in the air is preferred to perform the task in a simple way. On the other hand, text and signature writing is usually executed by resting the wrist and forearm, so the saturation increasing phenomenon is not observed.

To determine whether these differences are significant or not, a Wilcoxon rank sum test for equal medians is performed. The test performs a two-sided rank sum test of the hypothesis that two independent samples, represented by the pressure in session x and pressure in session y (Sx-Sy, $x, y \in [1,5]$), come from distributions with equal medians and returns the p-value from the test. The p-value represents the probability of observing the given result or one more extreme by chance if the null hypothesis ("medians are equal") is true. Small values of p cast doubt on the validity of the null hypothesis.

Table 2 shows the p-values obtained from pairwise comparisons between sessions. We conducted 10 comparisons for each task, considering the five sessions included in the database. As expected, we observed significant statistical differences when we compared extreme sessions (p-values<0.05). Moreover, as we noted earlier, the concentric loops and Archimedes spiral tasks showed high statistical differences across sessions 1 and 4. In other words, this analysis has confirmed our observation that more attention should be paid to tasks where fatigue has a greater impact on pressure signal saturation.

**Table 2.** Represents the p value obtained by Wilcoxon rank sum test when comparing different tasks and sessions.

| task | S1-S2 | S1-S3 | S1-S4 | S1-S5 | S2-S3 | S2-S4 | S2-S5 | S3-S4 | S3-S5 | S4-S5 |
|---|---|---|---|---|---|---|---|---|---|---|
| 1 | 0.642 | 0.943 | 0.162 | 0.977 | 0.600 | 0.056 | 0.667 | 0.187 | 0.860 | 0.178 |
| 2 | 0.329 | 0.160 | 0.070 | 0.104 | 0.639 | 0.328 | 0.404 | 0.630 | 0.734 | 0.845 |
| 3 | 0.463 | 0.642 | **0.025** | 0.130 | 0.806 | 0.113 | 0.468 | 0.079 | 0.346 | 0.462 |
| 4 | 0.967 | 0.463 | 0.614 | 0.399 | 0.533 | 0.727 | 0.463 | 0.889 | 0.866 | 0.806 |
| 5 | 0.296 | 0.208 | **0.041** | 0.208 | 0.865 | 0.394 | 0.874 | 0.399 | 0.977 | 0.545 |
| 6 | 0.920 | 0.946 | 0.659 | 0.795 | 1.000 | 0.528 | 0.672 | 0.549 | 0.672 | 0.878 |
| 7 | 0.973 | 0.729 | 0.920 | 0.624 | 0.82 | 0.946 | 0.599 | 0.682 | 0.346 | 0.599 |
| 8 | 0.698 | 0.757 | 0.441 | 0.419 | 0.898 | 0.806 | 0.726 | 0.624 | 0.599 | 0.946 |
| 9 | 0.359 | 0.076 | 0.575 | 0.674 | 0.348 | 0.727 | 0.587 | 0.180 | 0.166 | 0.869 |

## 4      Conclusions

In summary, this paper has introduced a novel feature for handwriting analysis, namely the percentage of saturated samples in a handwriting task. Our experimental findings demonstrate that this feature has discriminatory power in detecting fatigue during tasks that do not involve wrist resting. We believe that this feature could be further explored in different contexts and applications for automatic classification purposes. Overall, our work contributes to the development of more effective and accurate methods for analyzing handwriting data. In future studies, we intend to investigate this effect in other contexts and applications for the purpose of automatic classification, based on different databases.

## Acknowledgements

This work has been supported by a collaboration between the MINECO Spanish grants number PID2020-113242RB-I00 and PID2019-109099RB-C41